\documentclass[preprint,pre,aps,epsfig]{revtex4}
\newcommand{\BE}{\begin{equation}}
\newcommand{\EE}{\end{equation}}

\def\bq{\begin{equation}}
\def\eq{\end{equation}}

\begin{document}


\title{Fluctuation-driven directed transport in the presence \\of L{\' e}vy flights}

\author{D. del-Castillo-Negrete}
\email{delcastillod@ornl.gov}
\affiliation{Oak Ridge National Laboratory \\ Oak Ridge TN, 37831-8071}

\author{V.~Yu. Gonchar, A.~V. Chechkin}
\affiliation{Institute for Theoretical Physics NSC KIPT \\ Akademicheskaya st.1, Kharkov 61108
Ukraine}


\begin{abstract}
Numerical evidence of  directed transport driven by symmetric 
L{\' e}vy noise  in time-independent ratchet potentials in the absence of an
external tilting force is presented.  The results are based on the numerical solution of the fractional
Fokker-Planck equation in a periodic potential and the corresponding 
Langevin equation with L{\' e}vy noise. The L{\' e}vy noise drives the system out of thermodynamic
equilibrium and an up-hill net current is generated. For small values of the noise intensity there is an
optimal value of the  L{\' e}vy noise index yielding the maximum current. The direction and magnitude of
the current can be manipulated by changing the L{\' e}vy noise asymmetry  and the potential asymmetry. 

\end{abstract}

\maketitle



Fluctuation-driven transport is a topic of considerable theoretical and practical
importance. Its simplest realization is the
Brownian motion in which fluctuations give rise to diffusion.  In the presence of external forces,
fluctuations can have a highly nontrivial effect. When inertia is neglected, this problem can be
described with an stochastic,  overdamped Langevin equations of the form
\bq
\label{langevine}
\dot{x}= -\partial_x V + F+\xi(t)\, ,
\eq
where  $V$ is the potential, $F$ is an external  tilting force, and $\xi$ represents the fluctualtions.  Of
particular interest is the case of periodic potentials, $V(x+L)=V(x)$, lacking reflection symmetry. These
potentials, for which there is not an
$x_0$ such that 
$V(-x) = V(x+x_0)$, are known as ratchets (see Fig.~1).  The spatial symmetry breaking implies  a
preferred direction along which the restoring force on a particle at a potential minima is weaker.
Because of this, it is  tempting to think that in the presence of fluctuations, particles will tend to drift 
towards the weak part of the potential giving rise to a  ``down-hill" unidirectional net transport. 
However, it can be shown that, consistent with the second law of thermodynamics,  such directed
transport can not take place in the case of fluctuations in thermodynamic equilibrium (i.e., for $\xi$
corresponding to thermal Gaussian noise), $F=0$, and time independent potentials, $V=V(x)$.  
An appealing  discussion that illustrates this issue in a ratchet and pawl mechanical device can be found
in Ref.\cite{feynman}. Nevertheless, in the presence of {\em non-equilibrium}
perturbations, fluctuations can give rise to net directed transport in ratchet potentials. The realization
and exploitation of  this idea has generated a great amount of interest in the non-equilibrium statistical
mechanics of ``Brownian motors" spanning more than a decade, see for example Ref.~ \cite{reimann} and
references therein.   A trivial way to generate directed transport is by imposing biased perturbations,
e.g. a constant $F$.  However, what is highly nontrivial is the appearance  of directed currents in the
presence of  {\em symmetric},  unbiased, i.e. zero-mean, non-equilibrium perturbations.   Referring
to Eq.~(\ref{langevine}), non-equilibrium perturbations can be introduced in $F$, $V$ and $\xi$, and
ratchet models can be classified depending on how this is done. For example, in ``pulsating ratchets"
$V=V(x,f(t))$, in
``tilting ratchets" a time dependent  zero-mean tilting force $F=F(t)$ is assumed, and in ``temperature
ratchets" the  noise strength $T$ in $<\xi(t) \xi(s)>=2 \eta k_BT(t) \delta(t-s)$ is a function of time.
A less studied class of ratchets, which is the main focus of the present work,
incorporate non-equilibrium fluctuations by assuming non-Gaussian noises. 
Some examples include ratchet transport driven by symmetric Poissonian white shot noise with
exponential distributed amplitudes \cite{polish_hanggi}, and the use of colored non-Gaussian noise
\cite{wio}. 

In this letter we consider directed transport in ratchet
potentials driven by L{\' e}vy noise which gives rise to  anomalously large particle
displacements known as L{\' e}vy flights. 
A L\'{e}vy  flight  process is  a non-Gaussian,
non-stationary random process whose increments are independent and distributed according to an
$\alpha$-stable L\'{e}vy probability distribution function \cite{taqqu}. $\alpha$-stable  distributions
play a very important role because, according to the generalized central limit theorem
\cite{gnedenko-kolmogorov}, they are the attractors of distributions of sums of random variables with
divergent second moments . Also, these distributions exhibit slowly decaying tails describing random
processes with large events, and their self-similarity properties make them useful in the  description
of scale free transport and random fractal processes. These properties explain the ubiquitousness of
Levy statistic in many areas of science, engineering, and economics \cite{sch}.
The role of L{\' e}vy flights in confining
potentials has been addressed in the literature before. Examples include the study of the 
decay properties of the probability distribution function, the study
of unimodal-multimodal bifurcations during relaxation \cite{bifurcations},  and the barrier crossing
Kramers problem \cite{kramers_1_2}. 
However, despite the widespread recognition of the importance  of L{\' e}vy processes, 
the effect of  L{\' e}vy  noise in ratchet potentials has not been addressed.

Beyond its intrinsic theoretical interest,   one of our
motivations for the study presented here  is the problem of non-diffusive
transport in  plasmas. In a recent letter \cite{florin}  a ratchet-type transport model was proposed to
describe non-diffusive impurity  transport in  magnetically confined fusion plasmas. In that work it was
assumed that  the fluctuating electrostatic potential is a stationary, homogeneous Gaussian process,
and thus in order to introduce a non-equilibrium fluctuation an additional time variation of the
two-point Eulerian correlation function had to be assumed. Although these assumptions are reasonable
and physical realizable, recent  experiments and numerical simulations have shown that there are cases
in which fluctuations in turbulent plasmas exhibit L{\' e}vy statistics. For example, L{\' e}vy statistics
has been  observed in electrostatic edge turbulence in tokamaks and stellarators
\cite{levy_plasmas}, and in numerical simulations of pressure-gradient driven plasma turbulence
\cite{dcn_2005}.  Accordingly, an open problem of interest is to explore the role of L{\' e}vy noise in
ratchet-type transport  of impurities.  
Our results  provide indirect evidence that in the presence of non-Gaussian L{\' e}vy
fluctuations, a pinch effect might be present in the case of static potentials.  The ideas presented here
might also have an impact  on ratchet transport in biological systems and condense matter
in general  where there is evidence of non-Gaussian fluctuation phenomena. 

Our methodology consists of two complimentary approaches based on the Langevin equation
(\ref{langevine}) driven by L{\' e}vy noise, and in the solution of the corresponding space-Fractional Fokker
Planck (FFP) equation. For a discusion on the equivalence of the two formulations see Ref.~\cite{dubkov_etal_2005}.
Although the numerical methods for the solution of fractional  differential
equations is a rapidly developing field, the integration of the FFP equation with
fractional derivatives in space and periodic potentials has not been  treated before.  
In Ref.~\cite{heinsalu_etal_2006} a numerical study was presented of the FFP
equation in a tilted periodic potential in the presence of subdiffusion. The key difference between that work and our contribution
is that while  Ref.~\cite{heinsalu_etal_2006} deals with subdiffusive processes caused by non-Markovian memory effects modeled with
fractional derivatives in time, here we consider  superdiffusive processes caused by Levy flights modeled with fractional derivatives in
space.   The ratchet potential is given by
\bq
\label{potential}
V=V_0\left\{ 
\begin{array}{ll}
1- \cos\left[ \pi x/a_1 \right]\, , & \mbox{ $0 \leq x < a_1$} \\
1+ \cos\left[ \pi (x-a_1)/a_2 \right]\, , & \mbox{ $a_1 \leq x < L$} \, ,\\
\end {array}
\right.
\eq
where $V_0$ is the  amplitude, $L=a_1+a_2$ is the period, $V(x+L)=V(x)$, and $A=(a_1-a_2)/L$
is the asymmetry parameter. In all the calculations presented here, $V_0=L=1$. 
 Compared with the potential $V=V_0[ \sin(2 \pi x/L) +  0.25 \sin(4 \pi x/L)]$
typically used in the literature, e.g. Ref~\cite{reimann}, Eq.~(\ref{potential}) offers 
the advantage of an easier control of
the spatial asymmetry. As shown in Fig.~1, for $A=-0.274$  both  potentials
coincide.  
In the Langevin approach we assume that the noise satisfies $L(\Delta t) = \int_t^{t+\Delta t}
\xi(t') dt'$ where $L$ is a L{\' e}vy flight  process with characteristic function,
$\hat{p}_L(k,\Delta t)=\int_{-\infty}^\infty e^{i k x} p_L (x,\Delta t) d x$ given by  \cite{taqqu}
\bq
\label{levy_characteristic}
\hat{p}_L(k,\Delta t)=\exp \left\{
- \Delta t \chi |k|^\alpha \left[ 1 - i {\rm sign} (k) \beta \tan (\pi \alpha/2) \right] \right \}\, ,
\eq
where $\alpha$ is the L{\' e}vy index, $\beta$ is the skewness parameter, and $\chi$ is the L{\' e}vy
noise intensity.  Unless explicitly indicated, we will consider $\chi=1/2$.
Since for the ratchet problem the mean (first
moment) of the displacement must be finite,  the range for admissible $\alpha$ is $1 < \alpha \leq 2$, 
whereas
$-1\leq \beta \leq 1$.  The time step used in Langevin simulation was $\Delta t=10^{-3}$.  Details of the numerical scheme
can be found in Ref.~\cite{cheplas}.  For $\alpha=2$, Eq.~(\ref{levy_characteristic}) reduces to a
Gaussian corresponding to a diffusive (Brownian) process. As expected, in this case the problem reduces
to the Smoluchowski-Feynman ratchet which, as the insert in Fig.~1 shows, only exhibits a current in
the trivial case of a biased constant force $F$. 

One of the main obstacles in the numerical integration of equations containing fractional derivatives in
space is the fact that the left, $_aD_x^\alpha$, and the right, $_xD_b^\alpha$,
Riemann-Liouville fractional derivaties are in general singular at the $x=a$ and the $x=b$ boundaries
respectively \cite{podlu}. 
Here  we circumvent this problem by regularazing the fractional space derivatives using the
Caputo prescription, 
$^c_aD_x^\alpha P =  _aD_x^\alpha\left[ P(x)-P(a)-P'(a) (x-a) \right]$ and   $^c_xD_b^\alpha P =
_xD_b^\alpha\left[ P(x)-P(b)+P'(b) (b-x ) \right]$, and write the FFP equation as
\bq
\label{FFP}
\partial_t P = \partial_x \left[ P \partial_x V_{eff} \right] + \chi \left [ l ^c_aD_x^\alpha + 
r ^c_xD_b^\alpha  \right] P \, ,
\eq
where $V_{eff}=V(x) -F x$, and 
\bq
\label{a_13_2}
_{a}^cD_x^\alpha P =\frac{1}{\Gamma(2-\alpha)} \, \, \int_{a}^x\,  
\frac{P''(y)}{\left(x-y\right)^{\alpha-1}}  \,d y\,  ,
\eq
for $1 < \alpha < 2$, with  $P''=\partial_y^2 P(y,t)$. The expression for
$^c_xD_b^\alpha P$ follows from Eq.~(\ref{a_13_2}) by changing the integration limits to $\int_x^b$ and
interchanging $x$ and $y$ in the denominator. 
The  factors
$l= - (1+\beta)/[2 \cos(\alpha \pi /2)]$ and $r= - (1-\beta)/[2 \cos(\alpha \pi /2)]$ determine the
relative weight of the left and the right fractional derivatives. The
order of the fractional operators $\alpha$, the asymmetry parameter $\beta$,  and the diffusivity $\chi$
are given by the index, the skewness,  and the scale factor of the characteristic function of the Langevin
L{\' e}vy noise in Eq.~(\ref{levy_characteristic}).  
For the numerical integration of Eq.~(\ref{FFP}) we used a finite difference  scheme based on the 
Grunwald-Letnikov representation of the regularized fractional operators. 
Further details of the method can be found in Ref.~\cite{DiegoPhysPlas}. 
The integration domain, $x\in (a,b)=(-10,10)$,  covered twenty periods of the potential. We used a grid
size  $\Delta x=0.02$ and an integration time step $\Delta t=1.8 \times 10^{-3}$. The boundary
conditions were $P(a)=P(b)=0$, and the initial  condition was a higly peaked normalized Gaussian
distribution centered at
$x=0$. 

Figure~2 shows typical trajectories obtained from the Langevin equation for different L{\' e}vy noise 
indices $\alpha$ and degrees of potential asymmetry, $A$. Panel (a) illustrates the absence of directed
current in the Gaussian case, whereas  panel (b) shows a  positive bias  in the presence of
symmetric L{\' e}vy flights. As expected,  the L{\' e}vy noise gives rise to large  particle jumps
encompassing several periods.  Conclusive evidence of the existence of a  net directed current due to
symmetric L{\' e}vy flights is provide in Fig.~3.  Consistent with Fig.~2(b), a net positive current is
observed.  Note that, as in the case of Poissonian noise
\cite{polish_hanggi},  ``up-hill"  transport, i.e. transport the direction of the larger potential gradient, is
observed.  The dependence of the steady state current on the asymmetry parameter $A$
for $F=0$ is illustrated in Fig.~4. Excellent agreement between the Langevin and the
Fokker-Planck results is observed.  For a fixed value of the L{\' e}vy index $\alpha$, the current is a
monotonically decreasing function of the asymmetry and,  as expected, for fixed  $A$ the current
becomes smaller as
$\alpha$ approaches the Gaussian limit $\alpha=2$. Consistent with the symmetry $\partial_x
V(x,A)=-\partial_x V(L-x,-A)$, the curves  exhibit odd symmetry with respect the origin.  
Figure~5 shows the dependence of the  steady state current on the  L{\' e}vy index $\alpha$ in the
 $\beta=0$ symmetric case.  For large  $\chi$, the current is a monotonically
decreasing function of $\alpha$.  However, for small $\chi$ the relation is non-monotonic and,
depending on $A$, there is an optimal value of $\alpha$ that yields the maximum current. It is also
instructive to explore the variation of the current with $\alpha$ fixing the
scale factor $\sigma=\chi^{1/\alpha}$ of the L{\' e}vy distributions in Eq.({\ref{levy_characteristic}). In
this case the dependence seems to be always monotonic.  A problem of interest in ratchet transport is
the control of the magnitude and direction of the current. Figure~6 shows how this can be
accomplished  by manipulating the  L{\' e}vy noise asymmetry $\beta$ and the potential  asymmetry $A$.
Consistent with the Langevin simulation in Fig.~2(c), in Fig.~6(a) there is a net current with
$A=0$ and $\beta=0.25$.  Comparing with Fig.~3, for which $A=-0.274$ and $\beta=0$, it is observed
that the noise asymmetric can mimic the potential asymmetry.   As Fig.~6(b) shows, for a  given
ratchet potential, it is possible to find a compensating value of
$\beta$ for which the current vanishes. Moreover, as 
Fig.~6(c) shows, a further increase of the L{\' e}vy noise skewness leads to a current reversal, a result
consistent with the corresponding Langevin simulation shown in Fig.~2~(d).  

Summarizing. In this letter we have presented numerical evidence of  directed
transport driven by symmetric L{\' e}vy noise in time-independent ratchet potentials in the absence of
an external tilting force.  In the limit $\alpha=2$, the noise becomes  Gaussian, the fluctuations
are in  thermodynamic equilibrium,  and  consistent with the second law of thermodynamics the current
vanishes. However, for $\alpha \neq 2$, the L{\' e}vy noise drives the system out of thermodynamic
equilibrium and an up-hill  net current is generated. For small values of $\chi$ and a fixed potential
asymmetry, there is an optimal value of
$\alpha$ yielding the maximum current. The direction and magnitude of the current can be
manipulated by changing the L{\' e}vy noise asymmetry  and the potential asymmetry. 
As an application, we conjecture that a recently  proposed ratchet pinch mechanism in
magnetically confined plasmas might be present even in the case of  static
electrostatic fluctuations.

D. del-Castillo-Negrete acknowledge support from the  Oak Ridge National Laboratory, managed by
UT-Battelle, LLC, for the U.S. Department of Energy under contract DE-AC05-00OR22725. A. Chechkin
thanks the Slovak Academy Information Agency (SAIA) for financial support. 


\pagebreak
\newcounter{figlist}
\subsection*{FIGURE CAPTIONS}
\begin{list}
{FIG.~\arabic{figlist}.}{\usecounter{figlist}
                    \setlength{\labelwidth}{.55in}
                    \setlength{\leftmargin}{.55in}}

\item
Effective ratchet potential $V_{eff}=V(x) - Fx$  for $F=-1$ according to  Eq.~(2), 
with $A = -0.274$ (dashed line), $A=0.2$ (dotted line), and $A= -0.5$ (dashed-dotted line).
The solid line corresponds to the ratchet potential discussed in  Ref.~\cite{reimann}.
(see Fig.2.4) . The inset shows the well-known dependence on $F$ in the $\alpha=2$ Gaussian
case according to the Langevin simulations (dots) with $A = -0.274$, and the  result reported in 
Ref.~\cite{reimann} (solid line).

\item
Typical trajectories according to the numerical solution of the overdamped Langevin
Eq.(\ref{langevine}) with $L=1$, $\chi=0.5$ and: (a) $\alpha=2$  (Gaussian), $A=-0.274$;   (b)
$\alpha=1.50$, $\beta=0$, $A=-0.274$;  (c) $\alpha=1.50$, $\beta=0.25$,  $A=0$; (d)  $\alpha=1.50$
$\beta=-0.5$, $A=-0.274$.

\item 
Evidence of ratchet current in the Fractional Fokker-Planck model with $\alpha=1.5$,  $\beta=0$,
$\chi=0.5$, $F=0$ and $A=-0.274$. The solid line in the main figure shows the mean $<x>$ with respect
to $x = 0$ which folows a linear scaling (dashed line) corresponding to a current $<{\dot x}>=0.145$. 
The inset at the top shows the final and initial (dashed line) probability distribution function
superimposed with the ratchet potential in Eq.~(\ref{potential}). The inset at the bottom shows the
profile of the probability distribution function at the final time. 


\item
Steady current  as a function of ratchet potential asymmetry $A$, for $\beta = 0$, $\chi = 0.5$ and  
L{\' e}vy indices $\alpha = 1.5$, $\alpha=1.75$ and  $\alpha=1.90$. The solid  line with dots  denote the
results according to the Langevin model and the circles and crosses the Fractional Fokker-Planck
results.  The inset shows the dependence of the current on $F$ for $A=0.2$ and $A=-0.2$. The
curve in the middle is the Gaussian case, which shows no dependence on the sign of $A$. The top
(bottom) curve is the  $\alpha=1.5$ L{\' e}vy result for $A=-0.2$ ($A=0.2$).  

\item
Steady state current  $<{\dot x}>$ versus L{\' e}vy index $\alpha$ for 
(a) $\chi=0.5$, (b) $\chi=0.05$, (c)  $\sigma=0.707$,  and (d)  $\sigma=0.0707$.
 Curves 1, 2, 3, and 4 correspond to 
$ A = -0.274$, $A=- 0.4$, $A=- 0.5$ and $A=- 0.6$ respectively.

\item
Dependence of ratchet current on L{\' e}vy noise asymmetry in the fractional Fokker-Planck equation.
Panels (a), (b) and (c) show the probability  distribution function for $\alpha=1.5$ and different values
of $\beta$, and $A$. The fourth panel shows the
corresponding time evolution of the mean $<x>$.

\end{list}


\end{document}